\documentstyle[psfig,times]{mn}
\begin{document}
\hsize=6truein

\renewcommand{\thefootnote}{\fnsymbol{footnote}}

\title[]{$\beta-$model and cooling flows in X--ray clusters of galaxies}

\author[]
{\parbox[]{6.in} {Stefano Ettori \\
\footnotesize
Institute of Astronomy, Madingley Road, Cambridge CB3 0HA \\
settori@ast.cam.ac.uk
}}                                            
\date{}
\maketitle

\begin{abstract}
The spatial emission from the core of cooling flow clusters of galaxies
is inadequately described by a $\beta-$model 
(Cavaliere \& Fusco-Femiano 1976). 
Spectrally, the central region of these clusters are well approximated
with a two-temperature model, where the inner temperature represents the 
multiphase status of the core and the outer temperature is a measure of
the ambient gas temperature.
Following this observational evidence,
I extend the use of the $\beta-$model to a two-phase gas emission,
where the two components coexist within
a boundary radius $r_{\rm cool}$ and the ambient gas alone fills
the volume shell at radius above $r_{\rm cool}$. 
This simple model still provides 
an analytic expression for the total surface brightness profile:
\[ S(b) = S_{\rm cool}(0) \left[1-\left(\frac{b}{r_{\rm
cool}}\right)^2\right]^{0.5+3\beta_{\rm cool}} + S_{\rm amb}(0)
\left[1+\left(\frac{b}{r_{\rm c}}\right)^2\right]^{0.5-3\beta_{\rm
amb}}. \]
(Note in the first term the different sign with respect to the standard
$\beta-$model).  
Based upon a physically meaningful model for the X--ray emission,
this formula can be used (i) to improve significantly
the modeling of the surface brightness profile of cooling flow clusters of 
galaxies when compared to the standard $\beta-$model results, 
(ii) to constrain properly the physical characteristics of the
intracluster plasma in the outskirts, like, e.g., the ambient gas temperature.
\end{abstract}


\section{INTRODUCTION} 

To constrain the physical parameters of extended X-ray sources
(e.g. groups and clusters of galaxies), the observed 
surface brightness can be either geometrically deprojected
or, more simply, fitted with a model obtained from an
assumed distribution of the gas density.

Given the hydrostatic equilibrium within the cluster, the gravitational
potential supports both the gas and the galaxies distribution.
If the latter is approximated via the King approximation (1962) to the
inner portions of an isothermal sphere (Lane-Emden equation in Binney \& 
Tremaine 1987), the gas density is then written as:
\begin{equation}
\rho_{\rm gas} = \rho_{\rm 0} (1+x^2)^{-3 \beta /2},
\label{eq:rho_b}
\end{equation} 
where $x= r/r_{\rm c}$ and $r_{\rm c}$ is the core radius of the
distribution.

The surface brightness profile observed at the projected radius $b$,
$S(b)$, is the projection on the sky of the plasma emissivity,
$\epsilon (r)$:
\begin{equation}
S(b) = \int_{b^2}^{\infty} \frac{\epsilon \ dr^2}
{\sqrt{r^2 - b^2}}.
\label{eq:sb} \end{equation}

The emissivity is equal to
\begin{equation}
\epsilon (r) = \Lambda(T_{\rm gas}) \ n_{\rm p}^2
\ {\rm erg} \ {\rm s}^{-1} \ {\rm cm}^{-3},
\label{eq:em} \end{equation}
where $n_{\rm p} = \rho_{\rm gas} / (2.21 \mu m_{\rm p} )$ is the
proton density and the cooling function, $\Lambda(T_{\rm gas})$,
depends upon the mechanism of the emission 
(mainly due to bremsstrahlung at $T_{\rm gas} > 2.5$ keV).

Assuming isothermality and a $\beta$-model for the gas density 
(eq.~\ref{eq:rho_b}),
the surface brightness profile has an analytic solution 
(eq.~3.196.2 in Gradshteyn and Ryzhik 1965):
\begin{eqnarray}
S(b) & = & n_0^2 r_{\rm c} \Lambda(T_{\rm gas}) 
B(3\beta -0.5, 0.5) \left[ 1 + \left(\frac{b}{r_{\rm c}}\right)^2
\right]^{0.5 -3 \beta} 
\nonumber \\
 & = & S_0 (1+x^2)^{0.5 -3 \beta},
\label{eq:beta}
\end{eqnarray}
where the validity of the beta function $B(a,b)$ puts the strict
constraint $3 \beta > 0.5$ and
the cooling function $\Lambda(T_{\rm gas})$ does not change radially.

The $\beta-$model (Cavaliere \& Fusco-Femiano 1976, 1978) provides a good
representation of the observed surface brightness and has the
advantage to easily constrain the gas density distribution.

Elsewhere (Ettori 2000), I consider the effect of the presence of a
temperature gradient in the estimate of the $\beta-$model parameters.
In this paper, I will focus on the deficiency of the $\beta-$model in 
modeling in a satisfactory way the central emission from cooling flow
clusters of galaxies.

The cooling flows (e.g. Sarazin 1988, Fabian 1994) result in an 
enhancement of the
gas density in the central region due to the high cooling efficiency in
the cluster core. Recently, there have been attempts to model this
excess in emission with a generic double $\beta-$model, i.e. the sum of
two components of surface brightness (Ikebe et al. 1996, Xu et al. 1998,
Mohr et al. 1999, Reiprich \& B\"ohringer 1999). 
The correlation between the presence 
of a cooling flow and the necessity for a second $\beta-$model is well
indicated from this figure: Peres et al. (1998) find that 40 per cent of a
flux limited sample of 55 clusters of galaxies has a deposition rate of
more than 100 $M_{\odot}$ yr$^{-1}$; this is the same percentage of the
clusters in the Mohr et al. sample that are better modeled with a double
$\beta-$model instead of a single one (18 out of 45 clusters). 

The double $\beta-$model as sum of two $S(b)$ in eq.~\ref{eq:beta}, 
however, is not physically meaningful. In fact, a single isothermal 
temperature is usually assumed for both different density 
components that, therefore, are not in equilibrium. Moreover, 
data with high spatial resolution do not show evidence of a second 
inner core radius.

I present in this work a simple geometrical and physical model
of the emission from cooling flow clusters of galaxies. This
model relies on recent spectral evidence that the cluster plasma
can be described as a gas with two phases, one related to the cooling gas and
the other to the ambient medium. Assuming that the extended intracluster
gas density, $n_{\rm gas}$, is well described by a $\beta-$model, 
I show in the following section that an analytic expression for $S(b)$ 
can be obtained to describe the surface brightness from 
cooling flow clusters of galaxies. In Section~3, I apply this model
to real data of clusters with or without cooling flows. 
This model allows to handle the emissivity due to each component.
I discuss the physical implications of this in Section~4.
In Section~5, I present some concluding remarks.

\section{The two--phases emission model}

Recent spectral analyses of cooling flow clusters of galaxies
(Allen et al. 2000, White 2000) have shown
how the spectral capabilities of the present instruments are unable to
resolve all the fine structures of a multiphase gas, allowing just a
modeling with a two-phase component, one that describes the emission from
the central cooling gas and the other that takes into account
the extended emission from the ambient medium.

\begin{figure}
\psfig{figure=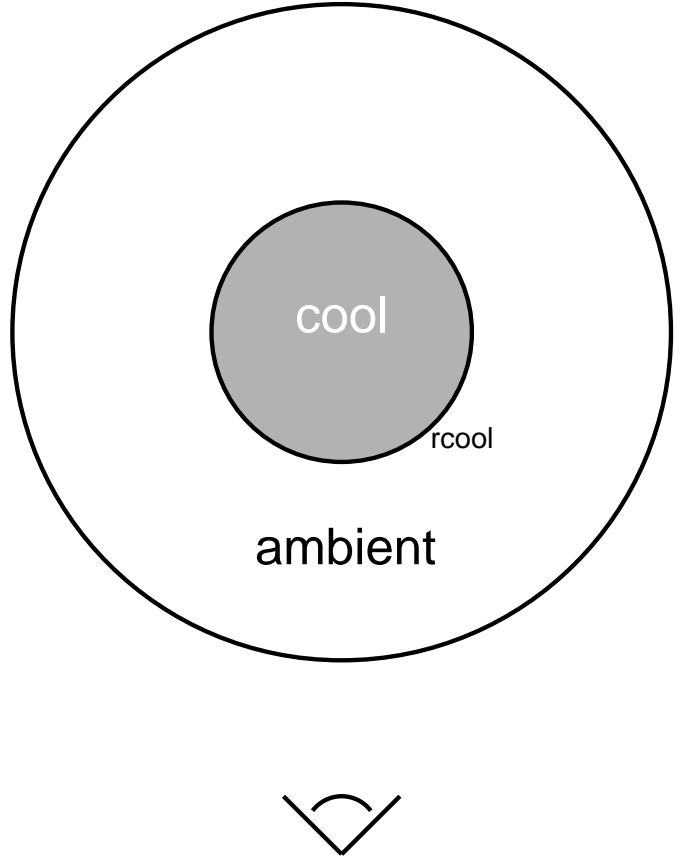,width=.5\textwidth}
\caption{Two-phases emission model. $r_{\rm cool}$ is the boundary
of the inner region and the only scale parameter used in the fit.
} \label{2phase} \end{figure}

These observational results provide us with a simple and natural model for
the total cluster emission: an inner cold phase confined within 
$r=r_{\rm cool}$ and overlapping the diffuse, ambient gas 
(see Fig.~\ref{2phase}). 

We assume that the two components coexist
within $r_{\rm cool}$, whereas only the ambient plasma 
fills the cluster volume shell at radius above $r_{\rm cool}$.
The total cluster emissivity is then $\epsilon(r,T) = \epsilon_{\rm cool}
+\epsilon_{\rm amb}$, where
\begin{equation}
\epsilon = \left\{ \begin{array}{l}
\Lambda(T_{\rm cool}) \ n_{\rm p, cool}(r)^2 +
\Lambda(T_{\rm amb}) \ n_{\rm p, amb}(r)^2, \ r < r_{\rm cool} \\
\Lambda(T_{\rm amb}) \ n_{\rm p, amb}(r)^2, \ r > r_{\rm cool} 
\end{array}
\right.
\end{equation}
from the definition in eq.~\ref{eq:em}.

This simple model provides an analytic expression for
the surface brightness profile defined in eq.~\ref{eq:sb}:
\begin{eqnarray}
S(b) & = & \int_{b^2}^{\infty} \frac{\epsilon \ dr^2} {\sqrt{r^2 - b^2}}
\nonumber \\
 & = & \int_{b^2}^{r_{\rm cool}^2} \frac{\epsilon_{\rm cool} \ dr^2}
{\sqrt{r^2 - b^2}} + \int_{b^2}^{\infty} \frac{\epsilon_{\rm amb} \ dr^2}
{\sqrt{r^2 - b^2}} \nonumber \\
 & = & S_{\rm cool}(b) + S_{\rm amb}(b)
\label{eq:sb1}
\end{eqnarray}
where the integration limits in $S_{\rm cool}(b)$ 
contains the boundary of the inner region at $r=r_{\rm cool}$.

Now, I integrate the emissivity along the line of sight.
$S_{\rm amb}(b)$ is still eq.~\ref{eq:beta}.
To integrate $S_{\rm cool}(b)$, one needs the assumption that the only
scale parameter of the gas density is the dimension of the cooling region,
$r_{\rm cool}$. Considering that we are in the regime $(r/r_{\rm cool}) <
1$, I can move the sign '--' from the exponent to the radix and
derive a $\beta-$model in the from of
\begin{equation}
n_{\rm p, cool} = n_{\rm 0, cool} \left[ 1 -
\left(\frac{r}{r_{\rm cool}}\right)^2 \right]^{1.5 \beta_{\rm cool} }.
\label{eq:np_cool}
\end{equation}
The behavior of this profile ensures that the gas density within the
cooling region has no other parameter scale than the dimension of the
region itself and $\epsilon_{\rm cool}$ goes to zero when $r
\rightarrow r_{\rm cool}$.

Then, I can integrate analytically $S_{\rm cool}(b)$ 
(eq.~3.196.3 in Gradshteyn and Ryzhik 1965):
\begin{eqnarray}
S_{\rm cool}(b) & = & S_{\rm cool}(0) \left[1-\left(\frac{b}{r_{\rm
cool}}\right)^2 \right]^{0.5+3\beta_{\rm cool}}, \nonumber \\
S_{\rm cool}(0) & = & n^2_{\rm 0, cool} r_{\rm cool} \Lambda(T_{\rm
cool}) B(0.5, 3\beta_{\rm cool}+1), 
\label{eq:s_cool}
\end{eqnarray}

In the equations above, $T_{\rm cool}$ and $T_{\rm amb}$ represent the two
gas temperatures corresponding to the cooling region and to the ambient of
the cluster, respectively.

\section{COMPARISON WITH THE DATA}

I have applied this model to observations of clusters of galaxies 
that can map the emission in regions well beyond the cluster core to 
disentangle the effect of the two components. Moreover, I have considered 
clusters with evidence of a large cooling flow and the Coma cluster
(ROR: rp800005n00, exposure time: 20.0 ksec, $z=0.0232$),
a well-known example of a no-cooling flow cluster.
In particular, I have analyzed, as described in Ettori and Fabian (1999), 
the {\it ROSAT} PSPC images of
A1795 (ROR: rp800105n00, exposure time: 33.3 ksec, $z=0.062$) and A2199
(ROR: rp800644n00, exposure time: 33.9 ksec, $z=0.030$), 
that have a deposition rate larger than 100
$M_{\odot}$ yr$^{-1}$ (Allen et al. 2000) and
present in the literature detailed spectral analyses 
of the {\it ASCA} dataset to be used as reference. 

\begin{figure*}
\hbox{
\psfig{figure=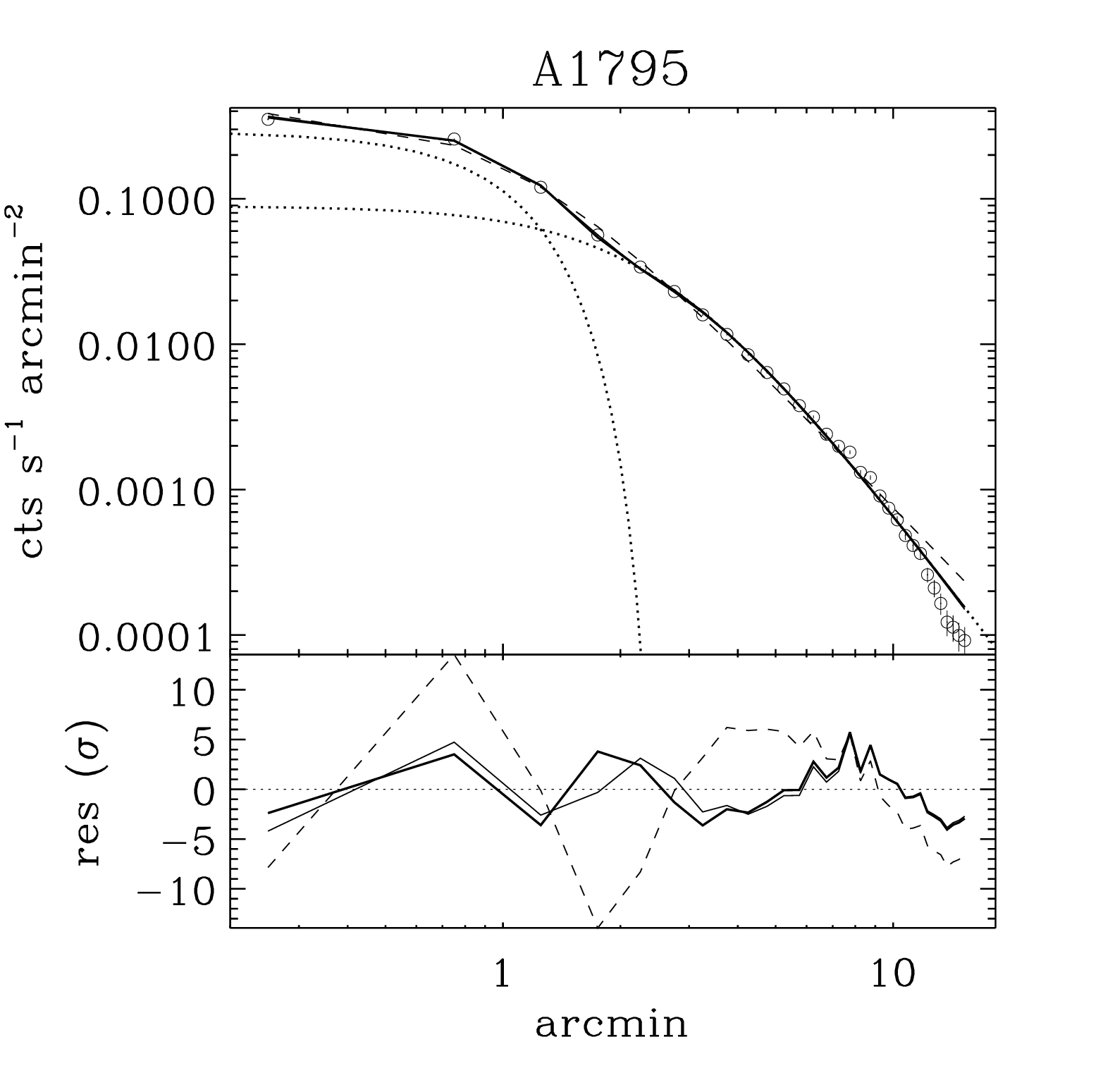,width=.5\textwidth}
\psfig{figure=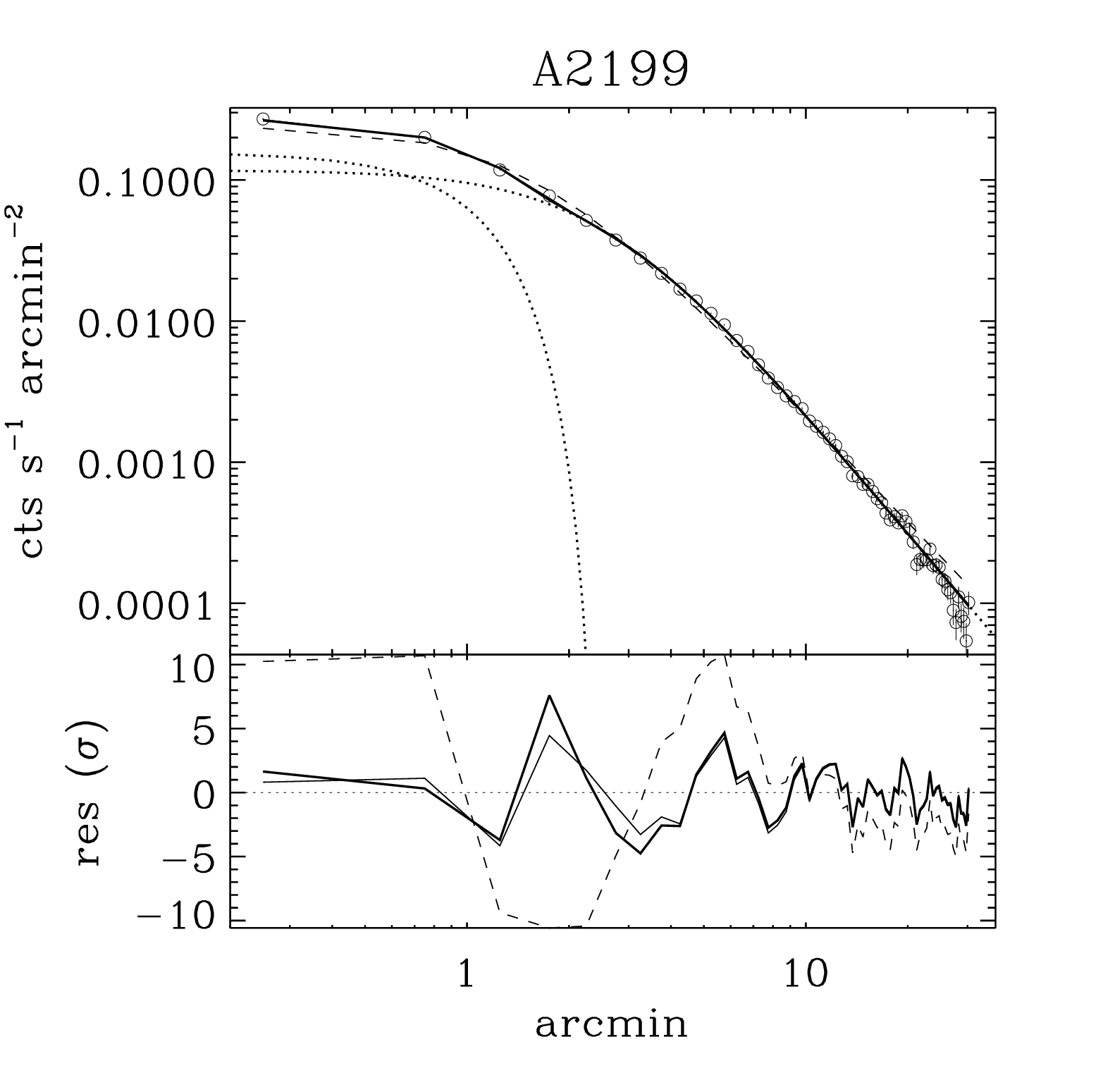,width=.5\textwidth}
}
\caption{The cluster surface brightness profile is here fitted with a
single $\beta-$model (dashed line) and a 2 $\beta-$model (solid line; the
thickest indicates the model with 5 parameters). 
The dotted lines show the two components of the 2 $\beta-$model 
with 5 parameters.
The panel below shows the residuals in unit of $\sigma$.
} \label{fig:data} \end{figure*}

In Table~1, I quote the results obtained by fitting the azimuthally
averaged profiles of the surface brightness between 0 and $R_{\rm out}$,
where the brightness value is larger than 3 times the uncertainty in that
radial bin. I perform the following fits (see Fig.~\ref{fig:data}):
(i) single $\beta-$model, (ii)
the double $\beta-$model presented here with 6 parameters $[S_{\rm
cool}(0), r_{\rm cool}, \beta_{\rm cool}, S_{\rm amb}(0), r_{\rm c},
\beta_{\rm amb}]$, (iii) the double $\beta-$model with 5 parameters, i.e.
fixing $r_{\rm c}=r_{\rm cool}$.

The decrease of the $\chi^2$ is significant where a cooling flow is present.
Where this decrease is not meaningful (or not present), like in Coma
cluster, the single 
$\beta-$model still represents a good description of the data.

Where a two-phase model provides a significantly better fit, 
I find that the F--test shows no statistical improvement
with a 6 parameters fit (cf. Table~1). 
Therefore, I use in the following considerations the
best fit results obtained with a 5 parameters fit,
i.e. $r_{\rm c}=r_{\rm cool}$.
This is not in contradiction with the present 
observational results. Allen (2000) quotes the cooling radii 
obtained from deprojection analysis of 30 cooling flow clusters images.
Ettori \& Fabian (1999) estimate the core radii for 23 of these clusters
using a single $\beta-$model over the radial range [0.1, 1] $r_{500}$,
the radius at which the mean cluster density is 500 the background value.
The distribution of the ratio, $r_{\rm c}/r_{\rm cool}$, has a median value 
of 1.33, an average of 1.61 and a standard deviation of 1.19, 
and can be considered consistent with $\sim 1$.
For the clusters in exam here, I measure a $r_{\rm c}/r_{\rm cool}$
ratio of 1.32$(\pm 0.24$, 90 per cent confidence level)  
and 0.84 $(\pm 0.15)$ for A1795 and A2199, respectively.
I remind, however, that I am using a different definition of $r_{\rm cool}$ 
than the one adopted in the standard spatial analysis: in the latter,
$r_{\rm cool}$ is the radius where the cluster cooling time first 
exceeds the Hubble time, whereas in this work $r_{\rm cool}$ defines
the boundary of the central cool phase of the gas.

It is worth to note that another version of a `5 parameters' fit, in which
$\beta_{\rm cool}$ is fixed equal to $\beta_{\rm amb}$ and $r_{\rm c}, 
r_{\rm cool}$ are left free to vary, provides a significantly worse
$\chi^2$ than the one obtained by using the `5 parameters' fit adopted
here. 

Finally, the $\chi^2$ obtained with the models above does not
vary significantly from the $\chi^2$ measured 
after the fit with other models which are not strictly based upon a 
physical framework, like the sum of two standard $\beta-$models
with 6 free parameters (e.g. Reiprich \& B\"ohringer 1999) or,
fixed the slope to a common value, with 5 free parameters 
(Mohr et al. 1999).

\begin{table*}
\caption{Best-fit results using the models discussed in the text. 
The core (cooling) radii are in $h^{-1}_{50}$ Mpc.
The values for the F--test represent the level of significance.
The symbols $2 \beta_5$ and $2 \beta_6$ indicate the two $\beta-$model with
5 and 6 parameters, respectively.
} \begin{tabular}{lcccccc} \hline
cluster & $R_{\rm out}$ & 1 $\beta$ & 2 $\beta$ -- 5 params & 2 $\beta$ -- 6 params &
\multicolumn{2}{c}{ F--test }\\
 & Mpc/$'$ & $r_{\rm c}, {\bf \chi^2}$(d.o.f.) & $r_{\rm cool}, {\bf \chi^2}$(d.o.f.) 
& $r_{\rm cool}, r_{\rm c}, {\bf \chi^2}$(d.o.f.) & $1 \beta \rightarrow 
2 \beta_5$ & $2 \beta_5 \rightarrow 2 \beta_6$ \\
 & & $\beta$ & $\beta_{\rm cool}, \beta_{\rm amb}$ &
$\beta_{\rm cool}, \beta_{\rm amb}$ & & \\
\hline
A1795 & 1.49/15.2 & $0.10\pm0.01, {\bf 40.9 (28)}$ & $0.26\pm0.01, 
  {\bf 8.6 (26)}$ & $0.64\pm0.15, 0.28\pm0.01, {\bf 8.2 (25)}$ \\
  & & $0.631\pm0.002$ & $1.801\pm0.126, 0.761\pm0.009$ &
  $4.459\pm1.012, 0.780\pm0.013$ & $>0.99$ & 0.45 \\ 
A2199 & 1.50/30.2 & $0.08\pm0.01, {\bf 23.1 (58)}$ & $0.13\pm0.01, 
  {\bf 4.7 (56)}$  & $0.74\pm0.06, 0.13\pm0.01, {\bf 3.6 (55)}$ \\
  & & $0.586\pm0.001$ & $1.635\pm0.058, 0.644\pm0.004$ & 
  $67.2\pm9.9, 0.650\pm0.005$ & $>0.99$ & 0.85 \\
Coma & 1.49/38.2 & $0.47\pm0.01, {\bf 8.2 (74)}$ & $0.51\pm0.01, 
  {\bf 6.5 (72)}$ & $33.67\pm3.05, 0.60\pm0.03, {\bf 8.6 (71)}$ \\
  & & $0.801\pm0.008$ & $9.071\pm1.532, 0.835\pm0.009$ & 
  $572.4\pm109.1, 1.376\pm0.116$ & 0.83 & 0.12 \\
\end{tabular}
\end{table*}

\section{DISCUSSION}

The use of this physically meaningful model allows us to directly handle
each of the two gas distributions, one that describes the profile
of the gas related to the cooling flow and the other that model 
the ambient gas.
As shown above, there is no statistical justification for using 
the 6 parameter fit.
Therefore, I consider hereafter the case $r_{\rm c} = r_{\rm cool}$.

We use in the following discussion the definition of the central gas
density for a cluster at redshift $z$, that is
\begin{equation}
n_{\rm p}(0) = \left[\frac{4\pi (1+z)^4 S(0)}{1.21 r_{\rm c} \Lambda(T) B}
\right]^{0.5}
\label{eq:n0}
\end{equation}
where $n_{\rm p}(0)$ is in cm$^{-3}$,
$r_{\rm c}$ is the core or cooling radius in cm, $S(0)$ is the
central surface brightness in cts s$^{-1}$ ster$^{-1}$, $B$ is the
proper beta function and the cooling function $\Lambda(T)$
is in unit of cts s$^{-1}$ cm$^5$.

Using the condition that the two phases have to be in pressure equilibrium,
I can now put constraints on their temperatures.
To do this, I consider the mean properties of each phase 
to handle integrated values instead of differential ones, because of 
the simple assumption that each phase is represented by a single
temperature that does not depend on the radius.
Therefore, each phase density, $\overline{n}(<r)$, is the integral of the 
radial density, $n(r)$, over the volume occupied from that phase 
(between 0 and $r_{\rm cool}$ for the inner phase; between 0 and 
$r = X r_{\rm cool}$ for the outer component)
divided by the integrated volume.
Then,
\begin{equation}
T_{\rm cool} = T_{\rm amb} \left( \frac{n_{\rm amb}}{n_{\rm cool}}
\right) = T_{\rm amb} \ g^{1/(2-\alpha)} \ I^{2/(2-\alpha)} = T_{\rm amb} \ f
\label{eq:t_law}
\end{equation}
where I have made use of the relation in eq.~\ref{eq:n0} 
$\Lambda(T) n_0^2 \sim S(0) / B(a,b)$, 
I have assumed $\Lambda(T) \sim T^{\alpha}$
($\alpha \approx 0.5$ for only bremsstrahlung emission observed by 
broad--band instruments), and I have defined 
\begin{eqnarray}
g & = & \frac{\Lambda(T_{\rm amb}) n_{\rm amb}(0)^2}
 {\Lambda(T_{\rm cool}) n_{\rm cool}(0)^2} = 
 \frac{S_{\rm amb}(0)}{S_{\rm cool}(0)} \frac{B_{\rm cool}}{B_{\rm amb}}
\nonumber \\
I & = & \frac{\overline{n_{\rm amb}}(<r)}{n_{\rm amb}(0)}
 \frac{n_{\rm cool}(0)}{\overline{n_{\rm cool}}(<r_{\rm cool})} \nonumber \\
 & = & \frac{ 1}{X^3} \frac{\int_0^X (1+x^2)^{-1.5 \beta_{\rm amb}} x^2 dx }
 {\int_0^1 (1-x^2)^{1.5 \beta_{\rm cool}} x^2 dx } \nonumber \\
 & = & \frac{I_X[1.5] / X^3}{ 0.5 B(1.5, 1.5\beta_{\rm cool} +1)} \nonumber \\
f & = & g^{1/(2-\alpha)} \ I^{2/(2-\alpha)},
\label{eq:param0}
\end{eqnarray}
with $\int_0^1 (1-x^2)^{1.5 \beta_{\rm cool}} x^2 dx =
B(1.5, 1.5 \beta_{\rm cool}+1)/2$ 
and $I_X[a] = \int_0^X (1+x^2)^{-a \beta_{\rm amb}} x^2 dx$. 

However, one generally measures a single emission-weighted temperature,
$\overline{T}$.
Given the considerations above, I can now disentangle the two components
(if any) using eq.~\ref{eq:t_law} in the following relation:
\begin{eqnarray}
\overline{T} & = & \frac{\int T \epsilon dV}{\int \epsilon dV} 
= \frac{\sum_{i=1,2} \int T_i \epsilon_i dV_i}
 {\sum_{i=1,2} \int \epsilon_i dV_i} 
\nonumber \\
 & = & T_{\rm amb} \frac{1 + f \ B(1.5, 3 \beta_{\rm cool}+1) \ 
(2 g \ I_X[3])^{-1}}{1 + B(1.5, 3 \beta_{\rm cool}+1) \ (2 g \ I_X[3])^{-1}}
\nonumber \\
 & = & T_{\rm amb} \ F,
\label{eq:f_maj}
\end{eqnarray}
where I still use the relation $\Lambda(T) n_0^2 \sim S(0) / B(a,b)$, 
calculate $\int_0^1 (1-x^2)^{3 \beta_{\rm cool}} x^2 dx = 
B(1.5, 3 \beta_{\rm cool}+1)/2$ and adopt the symbols $f, g, I_X$ defined
in eq.~\ref{eq:param0}. 

\begin{figure}
\psfig{figure=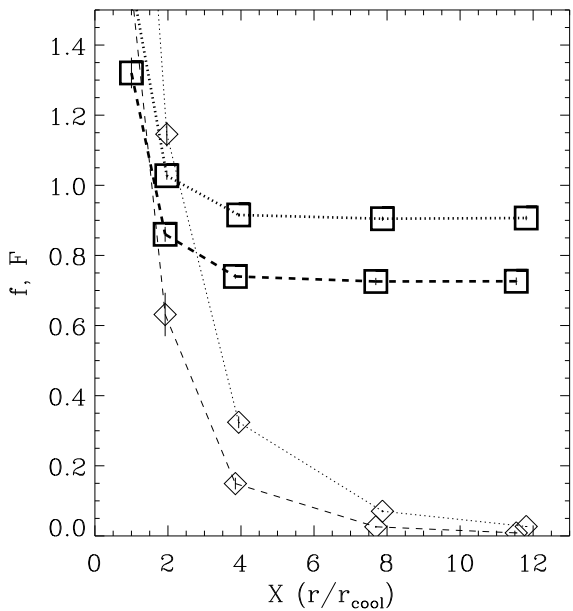,width=.5\textwidth}
\caption{Plot of the functions $f$ (diamonds) and $F$ (squares) vs.
$X$.  The dashed lines represent A1795, the dotted ones A2199.
The error-bars are overplotted to each point.
} \label{fig:f_par} \end{figure}

In the equations above, both the function $f$ and $F$ have to be smaller 
than 1 by definition. 
Their behaviour, however, depends strongly upon $X$, 
the radius in unit of $r_{\rm cool}$ up to where the outer phase extends
and can be represented with a single temperature.
Figure~\ref{fig:f_par} shows how the function $f$ and 
$F$ depend upon $X$: $f$ diminishes significantly due to the 
presence of $X^3$ in $I$ (eq.~\ref{eq:param0}), whereas $F$ 
converges quite rapidly (at $X \ga$ 4), providing a robust estimate on the 
$\overline{T} / T_{\rm amb}$ ratio.
Therefore, even if we are not able to constrain the ratio between 
the temperature of the two phases due to the uncertainty of the extension
of the outer component, we can assess the ambient temperature,
$T_{\rm amb}$, in a cooling flow cluster with an emission weighted 
temperature, $\overline{T}$, just using the azimuthally averaged 
surface brightness profile.

I show in Table~2 the constraints on $F = \overline{T} / T_{\rm amb}$ 
obtained from the spatial fit of the surface brightness profiles 
of the clusters in exam
and compare these values to the two-temperature spectral results
in Allen et al. (2000), Markevitch et al. (1998, 1999) and White (2000).
The agreement is remarkably good with the results of Markevitch and 
collaborators and White, which assume a two-phases gas for their 
spectral model in a way similar to the one I have adopted for the 
physical framework described above. (Note that Markevitch et al. 
measure the isothermal ambient temperature excising the cooling region, 
whereas White adds a cooling flow component in the spectral fit).
On the other hand, the disagreement with the results of 
Allen and collaborators can be explained with the more complex model
that they adopt, where an absorption intrinsic to the cluster is
combined with the cooling flow only.

\begin{table*}
\caption{Comparison between the ratio of the gas temperature obtained
spectrally from (i) a single temperature model and (ii) a two-temperature
(or isothermal + cooling flow components) model and the $F$ function in
eq.~\ref{eq:f_maj}.
The errors are at the 68.3 per cent confidence level.
In the observational data, these come from the propagation of the error 
on the temperature estimates. The error on the $F$ function is computed 
after 500 Monte Carlo fit of the original surface brightness profile.
References for the observational data: [1] Allen et al. (2000), [2]
Markevitch et al. (1998, 1999), [3] White (2000).  }
\begin{tabular}{lcc} \hline
cluster &  [1], [2], [3] & {\bf $F$: best-fit results} \\
\hline
A1795 & $0.870 \pm 0.023$, $0.769\pm0.065$, $0.799\pm0.057$ 
& ${\bf 0.726\pm0.010}$ \\
A2199 & $0.837 \pm 0.037$, $0.917\pm0.035$, $0.909\pm0.030$ 
& ${\bf 0.905\pm0.006}$ \\
\end{tabular}
\end{table*}

Several aspects of the cluster physical characteristics are affected
from the inclusion of a cooling flow in the modeling of the surface
brightness with a $\beta-$model.
With respect to the single $\beta-$model, one expects (i) excess in the
gas density in the cooling region, (ii) change in the $\beta_{\rm
amb}$ value, (iii) variations in the gas ambient temperature.

I present in Table~3 some of the more interesting physical quantities 
that can be evaluated with the equations above and given an 
emission-weighted temperature, $\overline{T}$ (from Allen et al. 2000:
$k\overline{T} = 5.40\pm0.05$ and $4.16\pm0.03$ keV for A1795 and A2199,
respectively).

For example, if one identifies the inner component with the cooling flow, 
a proper description of its gas distribution is now available through 
eq.~\ref{eq:np_cool}. The luminosity of the intracluster gas can then
be estimated without the contribution of the emission from the cool phase:
\begin{equation}
L_{\rm amb} =  \int 1.21 n^2_{\rm p}(r) \ \Lambda(T_{\rm amb}) \ 4\pi r^2 dr
\label{eq:lum}
\end{equation}
where the integral is computed upon the cluster volume and
$\Lambda(T)$ is here in erg s$^{-1}$ cm$^3$.
Using only the cluster surface brightness profile and a broad--band
emission-weighted gas temperature, and applying eq.~\ref{eq:f_maj} 
and \ref{eq:lum}, I will investigate in a forthcoming paper the effects 
on the clusters luminosity--temperature relation of the presence of
significant cooling flows
(see, e.g., the results from spectral analyses in Allen \& Fabian 1998, 
Markevitch 1998).


\begin{table*}
\caption{Ratios between the quantities discussed in the text and
estimated by using (i) a fit with a $1-\beta$ and $\overline{T}$ and (ii) 
a $2-\beta_5$ model and $T_{\rm amb}$, respectively. 
The cooling function is estimated using a MEKAL model 
(Kaastra 1992, Liedhal et al. 1995) in {\sc Xspec}
(version~10, Arnaud 1996).
The typical relative errors on the ratio of luminosities, gas and total masses
are of about 1 per cent and come 
from 500 Monte Carlo replications of the observed
surface brightness profile and emission-weighted temperature.
}
\begin{tabular}{lccccc} \hline
cluster &  $\Lambda(\overline{T})/\Lambda(T_{\rm amb})$ &
$\overline{L}/L_{\rm amb}$ & $\overline{M}_{\rm gas}/M_{\rm gas}$ & 
$\overline{M}_{\rm tot}/M_{\rm tot}$ \\
 & & 0.5, 1.0, 1.5 Mpc & 0.5, 1.0, 1.5 Mpc & 0.5, 1.0, 1.5 Mpc \\
\hline
A1795 & 
$0.87\pm0.01$ & 1.34, 1.18, 1.16 & 0.89, 0.92, 0.97 & 0.73, 0.64, 0.62  \\
A2199 & 
$0.96\pm0.01$ & 1.06, 1.04, 1.05 & 0.94, 0.98, 1.02 & 0.86, 0.84, 0.83  \\ 
 \\
\end{tabular}
\end{table*}

Appreciable corrections can also affect $M_{\rm gas}$,
$M_{\rm tot}$ and the terms of the so-called
$\beta-$problem (Mushotzky 1984, Edge \& Stewart 1991) due to the
variation of the $\beta$ value (for A1795 and A2199, $\beta_{\rm amb}$
increases by 20 and 10 per cent when compared to the $1-\beta$ model
fit results, respectively).
In the two-phases model described here, it is simple to calculate the gas
and the total mass:
$M_{\rm gas}$ is the integral of 
$\rho_{\rm gas} = \rho_{\rm cool} + \rho_{\rm amb} = 2.21 \mu m_{\rm p}
(n_{\rm p, cool} + n_{\rm p, amb})$
upon the cluster volume and, in particular,
\begin{equation}
M_{\rm gas}(r>r_{\rm cool}) = M_{\rm gas, cool} + \int_0^r \rho_{\rm amb}
\ 4\pi r^2 dr,
\end{equation}
where $M_{\rm gas, cool} = 2\pi \ \rho_{\rm 0, cool} \
r^3_{\rm cool} \ B(1.5, 1.5\beta_{\rm cool}+1) \approx 27 \ (3) 
\times 10^{11} M_{\odot}$ for A1795 (A2199) for an assumed $T_{\rm cool}
= (f/F) \ \overline{T} = 0.2 \overline{T}$;
the total gravitating mass is given by the application of the hydrostatic 
equilibrium,
\begin{equation}
M_{\rm tot} = \frac{r^2}{\mu m_{\rm p} G \rho_{\rm gas}} \frac{d(T_{\rm cool}
\rho_{\rm cool} + T_{\rm amb} \rho_{\rm amb})}{dr},
\end{equation}
and is proportional to $r_{\rm cool} \beta_{\rm amb} T_{\rm amb} 
x^3 (1+x^2)^{-1}$ at $r > r_{\rm cool}$.

\section{CONCLUSIONS}

I have presented a new analytic formula to model the total surface brightness
profile of clusters of galaxies where a two-phases intracluster gas can
be assumed. This scenario is consistent with the present results
of spectral analyses of the central regions of clusters that harbour a 
cooling flow.

The use of this formula allows to properly disentangle the contribution 
of the cooling flow to the cluster emissivity using only the spatial
distribution of the X-ray photons. After removing the contamination from the
cooling flow, I show how some relevant physical parameters are affected, 
like, for example, the ambient gas temperature (see Table~2).
In a forthcoming paper,
I will investigate the systematic changes in the temperature, luminosity and
mass (cf. Table~3) of a sample of clusters of galaxies and how these 
variations affect the relations among these quantities. 
 
\section*{ACKNOWLEDGEMENTS} 
I thank Anna, Sara and Carlo. 
I acknowledge the support of the Royal Society. Andy Fabian and 
David White are thanked for an useful reading of the manuscript, and 
the anonymous referee for comments which improved this work.

\end{document}